\documentstyle[preprint,prc,eqsecnum,aps,epsf]{revtex}

\tightenlines   %for single spacing in preprint

\input{psfig}

\begin{document}
\draft

\title{
Three-Body Scattering Below Breakup Threshold: An Approach
without using Partial Waves}

\author{W.~Schadow, Ch.~Elster}
\address{
Institute of Nuclear and Particle Physics,  and
Department of Physics, \\ Ohio University, Athens, OH 45701}

\author{W.~Gl\"ockle}
\address{
 Institute for Theoretical Physics II, Ruhr-University Bochum,
D-44780 Bochum, Germany.}

\date{July, 29, 1999}

\maketitle

\begin{abstract}
The Faddeev equation for three-body scattering below the three-body
 breakup
threshold is directly solved without employing a partial wave
decomposition. In the simplest form it is a three-dimensional integral
equation in four variables. From its solution the scattering amplitude
is obtained  as function of vector Jacobi momenta. Based on
Malfliet-Tjon type potentials  differential
and total cross sections are calculated. The numerical stability of
the algorithm is demonstrated and the properties of the  scattering
amplitude discussed.
\end{abstract}

\section{Introduction}

Traditionally three-nucleon scattering calculations are carried out by
solving Faddeev equations in a partial wave truncated basis. A partial
wave decomposition replaces the continuous angle variables by discrete
orbital angular momentum quantum numbers, and thus reduces the number
of continuous variables, which have to be discretized in a numerical
treatment. For low projectile energies the procedure of considering
orbital angular momentum components appears physically justified due
to arguments related to the centrifugal barrier. However, the
algebraic and algorithmic steps to be carried out in a partial wave
decomposition can be quite involved when solving the Faddeev
equations.  If one considers three-nucleon scattering at a few hundred
MeV projectile energy, the number of partial waves needed to achieve
convergence proliferates, and limitations with respect to
computational feasibility and accuracy are reached. The amplitudes
acquire stronger angular dependence, which is already visible in the
two-nucleon amplitudes, and their formation by an increasing number of
partial waves not only becomes more tedious but also less
informative. The method of partial wave decomposition looses its
physical transparency, and the direct use of angular variables becomes
more appealing.

It appears therefore natural to avoid a partial wave representation
completely and work directly with vector variables. This is common
practice in bound state calculations of few-nucleon systems based on
variational \cite{Arriaga95a} and Green's function Monte Carlo (GFMC)
methods \cite{Carlson87a,Carlson88a,Zabolitzki82a,Carlson99a}, which
are carried out in configuration space.

Our aim is to work directly with vector variables in the Faddeev
scheme in momentum space. In \cite{Elster98a} we demonstrated for
Malfliet-Tjon type potentials that the two-body Lippmann-Schwinger
equation can readily be solved in momentum space as function of vector
momenta.  At intermediate energies the strong forward peaking of the
$t$-matrix was quite easily achieved through the angular variable but
required relatively many partial wave contributions. The choice of
momentum vectors as adequate variables is also suggested by the
nucleon-nucleon (NN) force. Here the dependence on momentum vectors
can be rather simple, e.g. in the widely used one-boson-exchange
force, whereas the partial wave representation of this force leads to
rather complicated expressions \cite{Machleidt87}.

In \cite{Elster98b} we showed that the bound state Faddeev equation
has a rather transparent structure when formulated with vector
variables compared to the coupled set of two-dimensional integral
equations obtained in a partial wave decomposed form. Based on
Malfliet-Tjon type interactions it was demonstrated that the numerical
solution of the bound state equation using vector variables is
straightforward and numerically very accurate.

In this article we want to show that the solution of the three-body
scattering equation can also be obtained in a straightforward manner
when employing vector variables, i.e. magnitudes of momenta and angles
between the momentum vectors. In this work we concentrate on
scattering below the three-body breakup threshold. Though we avoid the
singularity of the free three-body propagator, we already encounter
the two-body fragmentation cut related to the pole in the two-body
$t$-matrix.  As a further simplification we neglect spin and iso-spin
degrees of freedom and treat three-boson scattering. The interactions
employed are of Yukawa type, and no separable approximations are
involved. The Faddeev equation for three identical bosons is
solved exactly as function of momentum vectors. To the best of our
knowledge this is the first time such an approach is carried out.

This article is organized as follows. Section II describes our choice
of momentum and angle variables for the unknown amplitude in the
Faddeev equation and the integral kernel of that equation.  The
calculation of the only scattering observable, the differential cross
section is also derived. In Section IIIa we discuss details of our
algorithms and numerical procedures and present our results. In
addition properties of the Faddeev amplitude are displayed. In
Section IV we discuss an alternative choice of variables and
demonstrate that this choice leads to the same result as the choice of
Section II. We conclude in Section V.

\section{Three-Body scattering equations}
\label{secscattereq}

We solve the Faddeev equations for three identical particles in the
form

\begin{equation}
\label{eqintt}
T | \phi \rangle = t P |  \phi \rangle + t G_0 P T |  \phi \rangle
\end{equation}

\noindent
where $t$ is the two-body $t$-matrix defined in the subsystem and the
operator $P$ is the sum of a cyclic and a anticyclic permutation of
three objects. The initial channel state $| \phi \rangle $ is composed
of a deuteron $|\varphi_{\rm d} \rangle$ and the momentum eigenstate
$|{\bf q_0} \rangle$ of the projectile nucleon. The free three-nucleon
propagator is given by $G_0 = (E - H_0 + i \epsilon) ^{-1}$, with $E$
being the total center of mass (c.m.) energy

\begin{equation}
E = \frac{3}{4m} q_0^2 + E_{\rm d}.
\end{equation}

\noindent
Here $E_{\rm d}$ is the binding energy of the two-body subsystem.
Since we work below the three-particle breakup threshold we only need to
consider the operator for elastic scattering

\begin{equation}
\label{eqintu}
U = P G_0^{-1} + P T.
\end{equation}

\noindent
From $U$ one obtains the differential cross section for elastic
scattering \cite{Gloeckle96a} as

\begin{equation}
\label{eqdcr}
\frac{d \sigma}{d \Omega}
  = \left ( \frac{2}{3} m \right )^2 (2 \pi)^4 \, \bigg |\langle
{\bf  q'} \varphi_{\rm d} | U | {\bf q}_0 \varphi_{\rm d} \rangle
 \bigg|^2,
\end{equation}

\noindent
where $| {\bf q}'| = | {\bf q}_0|$.
The total cross section is either obtained by integrating over
the angle variable

\begin{equation}
\label{eqdcrint}
\sigma_{\rm tot}^{\rm el} = \int d\Omega \, \frac{d \sigma^{\rm el}}{d
\Omega} = \left ( \frac{2}{3} m \right )^2 (2 \pi)^5
\int\limits_{-1}^{1} dx' \, \bigg |\langle q_0 x' \varphi_{\rm d} | U |
{\bf q}_0 \varphi_{\rm d} \rangle \bigg |^2
\end{equation}

\noindent
or via the optical theorem

\begin{equation}
\label{eqtotopt}
\sigma_{\rm tot} = - (2 \pi)^3 \, \frac{4 m }{3 q_0} \, {\rm
Im} \bigg ( \langle {\bf q}_0 \varphi_{\rm d} | U | {\bf q}_0
\varphi_{\rm d} \rangle \bigg ).
\end{equation}

In order to solve (\ref{eqintt}) we introduce the standard Jacobi
momenta ${\bf p}$, the relative momentum in the subsystem, and ${\bf
q}$ the relative momentum of the spectator with respect to the
subsystem.  With $|\phi \rangle = | {\bf q_0} \varphi_{\rm d} \rangle$
Eq.~(\ref{eqintt}) reads

\begin{equation}
\label{eqintt2}
\langle {\bf p} {\bf q} | T | {\bf q_0} \varphi_{\rm d} \rangle  =
\langle {\bf p} {\bf q} | t P |  {\bf q_0} \varphi_{\rm d} \rangle +
\langle {\bf p} {\bf q} | t G_0 P T | {\bf q_0} \varphi_{\rm d} \rangle .
\end{equation}

\noindent
The driving term of Eq (\ref{eqintt}) is given by

\begin{eqnarray}
\label{eqtp}
\langle {\bf p} {\bf q}|t \, P | \phi\rangle &=& \int d^3 q'\, d^3p'\,
d^3q'' \, d^3p'' \; \langle {\bf p} {\bf q}|t|{\bf p}'{\bf q}'\rangle
\, \langle {\bf p}'{\bf q}'|P|{\bf p}''{\bf q}'' \rangle \, \langle
{\bf p}''{\bf q}''|\phi\rangle \nonumber \\
&=& \int d^3 q'\, d^3p'\,
 d^3p'' \; \langle {\bf p} {\bf q}|t|{\bf p}'{\bf q}'\rangle
\, \langle {\bf p}'{\bf q}'|P|{\bf p}''{\bf q}_0 \rangle \, \langle
{\bf p}'' |\varphi_{\rm d} \rangle .
\end{eqnarray}

The momentum states are normalized according to $\langle {\bf p}'{\bf
q}'|{\bf p} {\bf q}\rangle = \delta^3 ({\bf p}'- {\bf p}) \, \delta^3
({\bf q}'-{\bf q})$. To evaluate the permutation operator $P = P_{12}
P_{23} + P_{13} P_{23}$ explicitly, the Jacobi coordinates in the
different subsystems $(12)$ and $(13)$ need to be expressed through
those defined in the subsystems $(23)$, which gives

\begin{eqnarray}
{\bf q}_1 &=& -{\bf p}_2 - \textstyle{\frac{1}{2}} {\bf q}_2
\nonumber \\ {\bf p}_1 &=&-\textstyle{\frac{1}{2}} {\bf p}_2 +
\textstyle{\frac{3}{4}} {\bf q}_2  \nonumber \\ {\bf q}_1 &=& {\bf p}_3
- \textstyle{\frac{1}{2}} {\bf q}_3  \nonumber \\ {\bf p}_1 &=& -
\textstyle{\frac{1}{2}} {\bf p}_3 - \textstyle{\frac{3}{4}} {\bf q}_3 .
\end{eqnarray}

\noindent
Then the permutation operator occuring in Eq.~(\ref{eqtp}) can be
evaluated as

\begin{eqnarray}
\langle {\bf p}'{\bf q}'|P|{\bf p}'' {\bf q}''\rangle &=&
  \langle {\bf p}'{\bf q}'|{\bf p}'' {\bf q}''\rangle_2 +
  \langle {\bf p}'{\bf q}'|{\bf p}'' {\bf q}''\rangle_3  \nonumber \\
 &=& \delta^3 ({\bf p}'+\textstyle{\frac{1}{2}}{\bf q}'+{\bf q}'') \,
     \delta^3 ({\bf p}''-{\bf q}'-\textstyle{\frac{1}{2}}{\bf q}'')
 \nonumber \\
 & &  + \,
     \delta^3 ({\bf p}'-\textstyle{\frac{1}{2}}{\bf q}'-{\bf q}'') \,
     \delta^3 ({\bf p}''+{\bf q}'+\textstyle{\frac{1}{2}}{\bf q}''),
 \label{eq:2.9}
\end{eqnarray}

\noindent
The indices 2 and 3 indicate the corresponding subsystem (for more
details see \cite{Elster98b}). Inserting this relation into
Eq.~(\ref{eqtp}) reduces the driving term to

\begin{equation}
\langle {\bf p} {\bf q}|t \, P | \phi\rangle
=  t_{\rm s} ({\bf p}, {\textstyle\frac{1}{2}} {\bf q} +
{\bf q}_0; E - {\textstyle{\frac{3}{4m}}} q^2) \, \varphi_{\rm d}
(|{\bf q} + {\textstyle\frac{1}{2}} {\bf q}_0| ) ,
\end{equation}

\noindent
were $t_{\rm s}({\bf p},{\bf q},E)$ is the symmetrized two-nucleon
$t$-matrix,

\begin{equation}
t_{\rm s}({\bf p},{\bf q},E)=t({\bf p},{\bf q},E)+t(-{\bf p},{\bf q},E) .
\end{equation}

Since we neglect spin, the deuteron consists only of an $S$ state, and
thus the wave function depends only on the magnitude of the momenta.

Carrying out a similar calculation for the integral term in
Eq.~(\ref{eqintt2}) leads to the explicit form of the  Faddeev
equation

\begin{eqnarray}
\label{eqintt3}
\lefteqn{ \langle {\bf p} {\bf q} |  T | {\bf q_0} \varphi_{\rm d}
\rangle =  t_{\rm s} ({\bf p}, {\textstyle\frac{1}{2}} {\bf q} +
{\bf q}_0; E - {\textstyle{\frac{3}{4m}}} q^2) \, \varphi_{\rm d}
(|{\bf q} + {\textstyle\frac{1}{2}} {\bf q}_0| ) } \nonumber \\ & &
\;\;\;\; + \int d^3 q''\,
\frac{ t_{\rm s} ({\bf p},
{\textstyle{\frac{1}{2}}} {\bf q} + {\bf q}''; E -
{\textstyle{\frac{3}{4m}}} q^2 )} {E - \frac{1}{m}(q^2 + {\bf
q} {\bf q}'' + q''^2) }\; \langle {\bf q} +
{\textstyle{\frac{1}{2}}} {\bf q}'', {\bf q}'' |  T | {\bf q}_0
\varphi_{\rm d} \rangle.
\end{eqnarray}

\noindent
The transition operator $T$  is needed for all values of ${\bf q}$.  Thus
one encounters the pole of the two-body $t$-matrix at the bound-state
energy $E_{\rm d}$. Extracting explicitly the residue by defining

\begin{equation}
\label{eqtmatres}
t_{\rm s} ({\bf p},{\bf q},E) = \frac{\hat t_{\rm s} ({\bf p},{\bf q},E)}{E
-E_{\rm d}}
\end{equation}
 and similarily for $T$, Eq. (\ref{eqintt3}) can be written as

\begin{eqnarray}
\label{eqinttvar}
\lefteqn{ \langle {\bf p} {\bf q} | \hat T | {\bf q_0} \varphi_{\rm d}
\rangle = \hat t_{\rm s} ({\bf p}, {\textstyle\frac{1}{2}} {\bf q} +
{\bf q}_0; E - {\textstyle{\frac{3}{4m}}} q^2) \, \varphi_{\rm d}
(|{\bf q} + {\textstyle\frac{1}{2}} {\bf q}_0 |) } \nonumber \\ & &
\;\;\;\; + \int d^3 q''\, \frac{ \hat t_{\rm s} ({\bf p},
{\textstyle{\frac{1}{2}}} {\bf q} + {\bf q}''; E -
{\textstyle{\frac{3}{4m}}} q^2 )}{E - \frac{1}{m}(q^2 + {\bf
q} \cdot {\bf q}'' + q''^2) } \; \frac{\langle {\bf q} +
{\textstyle{\frac{1}{2}}} {\bf q}'', {\bf q}'' | \hat T | {\bf q}_0
\varphi_{\rm d} \rangle}{E - \frac{3}{4 m} q''^2 - E_{\rm d} + i
\epsilon}.
\end{eqnarray}

\noindent
This expression  is the starting point for our numerical calculation of the
transition amplitude. Correspondingly the operator for elastic
scattering in Eq. (\ref{eqintu}) reads

\begin{eqnarray}
\label{eqintuvar}
\langle {\bf q}' \varphi_{\rm d} | U | {\bf q_0} \varphi_{\rm d} \rangle &
= & 2 \, \varphi_{\rm d} (\textstyle{\frac{1}{2}} {\bf q}' + {\bf q_0})
 \left (E -
\textstyle{\frac{1}{m}} ( q'^2 + {\bf q}' {\bf q}_0 + q_0^2)
\right) \varphi_{\rm d} ( {\bf q}' + \textstyle{\frac{1}{2}} {\bf q}_0)
\nonumber \\ & & + \, 2 \int d^3 q''
\varphi_{\rm d} (\textstyle{\frac{1}{2}} {\bf q}' + {\bf q}'')
\langle {\bf q}' + \textstyle{\frac{1}{2}} {\bf q}'', {\bf q}'' | T | {\bf q}_0
\varphi_{\rm d} \rangle.
\end{eqnarray}

The transition amplitude $\hat T$ as given in Eq.~(\ref{eqinttvar})
depends on the vector variables ${\bf q}_0$, ${\bf q}$, and ${\bf
p}$. Going to c.m. coordinates and choosing the $z$-axis in the  direction
of ${\bf q}_0$ we are left with five independent variables. Those are
the magnitudes of the vectors ${\bf q}$ and ${\bf p}$, their angles
with respect to the $z$-axis and the azimuthal angle $\varphi_{pq}$
between them. The vectors ${\bf q}_0$ and ${\bf q}$ define the $x$-$z$
plane in a cartesian coordinate system. In these variables the matrix
element for the transition amplitude can be written as $\langle p,
x_p, \cos \varphi_{pq}, q, x_q | T | {\bf q}_0 \varphi_{\rm d}
\rangle$ with

\begin{eqnarray}
\label{eqvars}
p &=& | {\bf p} |  \nonumber \\
q &=& | {\bf q} | \nonumber \\
x_p & = & \hat {\bf p} \cdot \hat {\bf q}_0 \nonumber \\
x_q &=& \hat {\bf q} \cdot \hat {\bf q}_0 \nonumber \\
\cos \varphi_{pq} & = & \cos \varphi ({\bf p}, {\bf q})
= \hat {\bf p}_{xy} \cdot  \hat {\bf q}_{xy}.
\end{eqnarray}

\noindent
The index $xy$ denotes the projection of the vectors into the
$x$-$y$ plane.
In order to obtain the matrix elements $\langle {\bf p} {\bf q} | \hat
 T| {\bf q}_0 \varphi_{\rm d} \rangle$, Eq.~(\ref{eqinttvar}) needs to be
solved. For the integration we choose  the $z$ axis parallel to
${\bf q}$. This implies that the azimuthal angle $\varphi$
between $({\bf q} + \textstyle{\frac{1}{2}} {\bf q''})$ and ${\bf
q''}$  for
$\langle {\bf q} + \textstyle{\frac{1}{2}} {\bf q''}, {\bf q}'' | \hat
 T| {\bf q}_0 \varphi_{\rm d} \rangle$ in the kernel of Eq.~(\ref{eqinttvar})
is zero, and thus  $\cos \varphi_{pq} = 1$ in Eq.~(\ref{eqvars}).
This also means that we only need to solve Eq.~(\ref{eqinttvar}) for
$\cos \varphi_{pq} = 1$, or in other words, that ${\bf p}$ lies in the
same plane that is spanned by ${\bf q}$ and ${\bf q}''$. From these
considerations follows that we only have to solve Eq.~(\ref{eqinttvar})
for ${\langle  p, x_p, 1,  q, x_q | \hat T | {\bf q_0}
\varphi_{\rm d} \rangle}$, i.e. for {\bf four} independent variables
instead of five, as it could be assumed from the considerations
proceeding Eq.~(\ref{eqvars}). Thus, for our calculations we arrive
from Eq.~(\ref{eqvars}) with $\cos \varphi_{pq} = 1$ at the following
two additional angle variables

\begin{eqnarray}
x_{pq} &=& \hat {\bf p} \cdot  \hat {\bf q} =  x_p x_q + \sqrt{1- x_p^2}
    \sqrt{1 -x_q^2} \cos \varphi_{pq}
=  x_p x_q + \sqrt{1- x_p^2} \sqrt{1 -x_q^2} \nonumber \\
x'' &= & \hat {\bf q} \cdot  \hat {\bf q}''.
\end{eqnarray}

\noindent
The angle variables $x_p$ and $x_q$ are already given in
Eq.~(\ref{eqvars}).  The momenta occurring in Eq.~(\ref{eqinttvar})
are now given explicitly as

\begin{equation}
\begin{array}{rclrcl}
|   {\textstyle{\frac{1}{2}}} {\bf q} + {\bf q}_0  | & = &
\sqrt{{\textstyle{\frac{1}{4}}} q^2 + q q_0 x_q +  q_0^2}
\;\;\;\; &
| {\textstyle{\frac{1}{2}}} {\bf q} +  {\bf q}'' | & = &
\sqrt{ {\textstyle{\frac{1}{4}}} q^2 + q q'' x'' +  q''^2} \\
|   { {\bf q} + \textstyle{\frac{1}{2}}} {\bf q}_0  | & = &
\sqrt{ q^2 + q q_0 x_q +  {\textstyle{\frac{1}{4}}} q_0^2}
&
|  {\bf q} + {\textstyle{\frac{1}{2}}} {\bf q}'' | & = &
\sqrt{ q^2 + q q'' x'' +  {\textstyle{\frac{1}{4}}} q''^2}.
\end{array}
\end{equation}

\noindent
Expressions for the remaining angle variables can be found in Appendix
\ref{anglest}. Inserting all variables into Eq.~(\ref{eqinttvar}), the final
expression for the transition amplitude reads

\begin{eqnarray}
\label{eqinttmomexpl}
\lefteqn{\langle  p, x_p, 1,  q, x_q | \hat T | {\bf q_0}
\varphi_{\rm d} \rangle} \\
  & =&  \hat t_{\rm
s} ( p, \sqrt{\textstyle\frac{1}{4}  q^2 + q q_0 x_q + q^2_0},
\frac{\frac{1}{2} q x_{pq} + q_0 x_p}{ \sqrt{\frac{1}{4} q^2 + q q_0 x_q
+ q_0^2 }}; E - {\textstyle{\frac{3}{4m}}} q^2) \;
  \varphi_{\rm d} (\sqrt{ q^2 + q q_0 x_q  +
 {\textstyle\frac{1}{4}}  q_0^2} ) \nonumber \\
& &  + \int\limits_{0}^{\infty} dq'' q''^2
\int\limits_{-1}^{1} dx'' \int\limits_{0}^{2 \pi} d\varphi''
\;  \hat t_{\rm
s} ( p, \sqrt{\textstyle\frac{1}{4}  q^2 + q q'' x'' + q''^2},
\frac{\frac{1}{2} q x_{pq} + q'' y_p}{ \sqrt{\frac{1}{4} q^2 + q q'' x''
+ q''^2 }}; E - {\textstyle{\frac{3}{4m}}} q^2) \nonumber \\
& & \,\,\, \times  \frac{1}{E - \frac{1}{m} (q^2 + q q'' x'' + q''^2)}
\frac{\langle \sqrt{q^2 + q q'' x'' + \frac{1}{4} q''^2},
\frac{ q x_q + \frac{1}{2} q'' y_{q_0}}{ \sqrt{ q^2 + q q'' x''
+ \frac{1}{4} q''^2 }}, 1, q'', y_{q_0}| \hat T | {\bf q_0}
\varphi_{\rm d} \rangle}{E - \frac{3}{4m} q''^2 - E_{\rm d} + i \epsilon}.
\nonumber
\end{eqnarray}

\noindent
This is a three-dimensional integral equation in four variables,
namely $p$, $x_p$, $q$, and $x_q$. The advantage of our choice of the
coordinate system is that the free propagator has a relatively simple
form, it depends only on the magnitude of momenta and one
angle. Though in the present work we stay with our calculations below
the breakup threshold, this particular form of the propagator will be
the most suited form for considering the solution of
Eq.~(\ref{eqinttmomexpl}) above.

The matrix elements of $\hat T$ provide input to the calculation of
the matrix elements $\langle {\bf q}' \varphi_{\rm d} | U | {\bf q_0}
\varphi_{\rm d} \rangle$ according to Eq.~(\ref{eqintuvar}).  In this
integration we choose  the $z$-axis parallel to ${\bf q}'$, so that there
is no azimuthal angle between $({\bf q}' + \textstyle{\frac{1}{2}}
{\bf q''})$ and ${\bf q''}$.  This specific choice ensures that we
only need $\hat T$ as function of four variables.  For the explicit
representation of $U$ the follwoing angle variables are are needed,
together with the magnitude of ${\bf q}'$ and ${\bf q}''$

\begin{eqnarray}
 x' &=&  \hat {\bf q}' \cdot \hat {\bf q}_0  = \cos \vartheta' \nonumber \\
 x'' &=  &  \hat {\bf q}' \cdot \hat {\bf q}'' = \cos \vartheta''
\end{eqnarray}

\noindent
and the momenta

\begin{eqnarray}
|   {\textstyle{\frac{1}{2}}} {\bf q}' + {\bf q}_0  | & = &
\sqrt{{\textstyle{\frac{1}{4}}} q'^2 + q' q_0 x' +  q_0^2}
 = q_0 \sqrt{\frac{5}{4} + x'} \nonumber \\
|   { {\bf q}' + \textstyle{\frac{1}{2}}} {\bf q}_0  | & = &
\sqrt{ q'^2 + q' q_0 x' +  {\textstyle{\frac{1}{4}}} q_0^2}
 = q_0 \sqrt{\frac{5}{4} + x'} \nonumber \\
| {\textstyle{\frac{1}{2}}} {\bf q}' +  {\bf q}'' | & = &
\sqrt{ {\textstyle{\frac{1}{4}}} q'^2 + q' q'' x'' +  q''^2} \nonumber \\
|  {\bf q}' + {\textstyle{\frac{1}{2}}} {\bf q}'' | & = &
\sqrt{ q'^2 + q' q'' x'' +  {\textstyle{\frac{1}{4}}} q''^2} .
\end{eqnarray}

\noindent
The explicit expressions for the remaining angles are calculated in
Appendix \ref{anglesu}. The final expression for the elastic  scattering
amplitude is then given by

\begin{eqnarray}
\label{eqintumomexpl}
\lefteqn{ \langle {\bf q'} \varphi_{\rm d} | U | {\bf q_0} \varphi_{\rm d}
\rangle  =  \langle q_0 x' \varphi_{\rm d} | U | {\bf q_0}
\varphi_{\rm d} \rangle } \nonumber \\ & = & 2 \, \varphi^2_{\rm d} (q_0
\sqrt{\textstyle{\frac{5}{4}} + x'}) \left (E -
\textstyle{\frac{q^2_0}{m}} (2 + x') \right ) \nonumber \\ & & + \, 2
\int\limits_{0}^{\infty} d q'' \, q''^2 \frac{1}{E - \frac{3}{4m}
q''^2 - E_{\rm d} + i \epsilon} \int\limits_{-1}^{1} dx''
\int\limits_{0}^{2 \pi} d \varphi'' \varphi_{\rm d} (
\sqrt{ \textstyle{\frac{1}{4}} q_0^2 + q_0
q'' x'' + q''^2 }) \nonumber \\ & & \;\;\;\;
\times \langle \sqrt{q_0^2 + q_0 q'' x'' + \textstyle{\frac{1}{4}}
q''^2}, \frac{q_0 x' + \frac{1}{2} q'' y}{\sqrt{q_0^2 + q_0 q'' x''+
\textstyle{\frac{1}{4}} q''^2}}, 1, q'' , y | \hat
T | {\bf q}_0 \varphi_{\rm d} \rangle.
\end{eqnarray}

\noindent
From $ \langle q_0 x' \varphi_{\rm d} | U | {\bf q_0}
\varphi_{\rm d} \rangle$ we obtain the differential cross section
according to  Eq.~(\ref{eqdcr}), and  the total cross section
via Eqs.~(\ref{eqdcrint}) and (\ref{eqtotopt}).

\section{Calculation of scattering observables}
\label{secscatterobs}

For our model calculations Yukawa interactions of
Malfliet-Tjon \cite{Malfliet69} type are used,

\begin{equation}
V({\bf p'},{\bf p})= \frac{1}{2\pi^2}\left( \frac{V_{\rm R}}{({\bf
      p'}- {\bf p})^2 + \mu_{\rm R}^2} - \frac{V_{\rm A}}{({\bf
      p'}-{\bf p})^2 + \mu_{\rm A}^2} \right).  \label{eq:3.1}
\end{equation}

\noindent
We study two different types of pairwise forces, a purely attractive
Yukawa interaction and a superposition of a short-ranged repulsive and
a long-ranged attractive Yukawa interaction.  It should be pointed out
that we calculate the potentials as functions of vector momenta and
thus define the interaction as a truly local force acting in all
partial wave.  The parameters are given in Table \ref{tabpotpar},
which also lists the corresponding deuteron binding energies.  The
parameters are chosen such that the deuteron binding energy is close
to the experimental one.  With these interactions we first solve the
Lippmann-Schwinger equation for the fully-off-shell two-nucleon
$t$-matrix directly as function of the vector variables as described
in detail in Ref.~\cite{Elster98a}.  The resulting $t$-matrix is then
symmetrized to get $t_{\rm s}( p', p, x;E-\frac{3}{4m}q^2)$. We would
like to point out that after having solved the Lippmann-Schwinger
equation on Gaussian grids for $p$, $p'$, and $x$, we solve the
integral equation again to obtain the $t$-matrix at points $x=\pm
1$. Thus, when solving Eq.~(\ref{eqinttvar}), we do not have to
extrapolate numerically to angle points $x$ of $t_{\rm s}( p', p,
x;E-\frac{3}{4m}q^2)$, which can very well be located outside the
upper or lower boundary of the Gaussian angle grid of the $t$-matrix.

The fully off-shell $t$-matrix, $t(p',p,x,E)$, is obtained for each
fixed energy on a symmetric momentum grid, which is divided as
$(0,p_1)\cup (p_1,p_{\rm max})$. The intervals contain NP1 and NP2
Gauss points, with typical values of NP1 = 40 and NP2 = 16
points. Typical values for the interval boundaries are $p_1 = 20$
fm$^{-1}$ and $p_{\rm max} = 60$ fm$^{-1}$.  For the angular
integration $x$ 32 Gauss points are sufficient.  Since the momentum
region which contributes to a solution of the two-body $t$-matrix is
quite different from the region of importance in a three-body
calculation, we map our solution for $t_{\rm s}$ onto a momentum grid
relevant for the three-body transition amplitude. This is done by
applying the Lippmann-Schwinger equation repeatedly. The $t$-matrix
$t_{\rm s}(p',p,x, \epsilon)$ is obtained at energies $\epsilon = E -
\frac{3}{4m} q^2$, exactly at the $q$ values needed in the three-body
transition amplitude of Eq.~(\ref{eqinttmomexpl}).  For extracting the
residue of the two-body $t$-matrix, Eq.~(\ref{eqtmatres}),
we represent $t_{\rm s}$ as

\begin{equation}
\label{eqtpole}
 (E- E_d) \, t_{\rm s}  \,
\raisebox{-2mm}
{$\stackrel{\longrightarrow}{\scriptscriptstyle{E \to E_{\rm d}}}$}
\, V | \varphi_{\rm d} \rangle \langle  \varphi_{\rm d} | V.
\end{equation}

\noindent
The Malfliet-Tjon type potentials support only an $s$-wave bound state,
and thus Eq.~(\ref{eqtpole}) reads explicitly

\begin{eqnarray}
\lefteqn{ (E- E_d) \, t_{\rm s}({\bf p}, {\bf p'})} \nonumber \\
& &  \,
\raisebox{-2mm}
{$\stackrel{\longrightarrow}{\scriptscriptstyle{E \to E_{\rm d}}}$}
\, \frac{1}{4 \pi} \left \{ \int\limits_0^\infty dp''
p''^2 \, V_0(p,p'') \varphi_{\rm d}(p'') \right \} \left \{ \int\limits
_0^\infty dp'' p''^2 \, V_0(p ',p'') \varphi_{\rm d}(p'') \right \},
\end{eqnarray}

\noindent
were $V_0$ is the $l = 0$ component of the potential.

In order to solve Eq.~(\ref{eqinttmomexpl}) we follow the iterative
procedure outlined in Ref. \cite{Kloet73}. The method consists of
first generating the Neumann series of Eq.~(\ref{eqinttmomexpl}) and
then summing up the series using the Pad{\'e} method
\cite{Baker70,Gloeckle83,Hueber96a}. We typically need to sum 15-18
terms to obtain a converged result. This is not surprising, since due
to the presence of the  three-body bound state the Neumann series itself diverges.

The $q'$-integration in Eq.~(\ref{eqinttmomexpl}) is cut off at a
value of $q_{\rm max} = 20$~fm$^{-1}$.  The integration interval is
divided into two parts, $(0,q_1)\cup (q_1,q_{\rm max})$, in which we
use Gaussian quadrature with NQ1 and NQ2 points, respectively. The
value for $q_1$ is chosen to be 5~fm$^{-1}$. Typical values for NQ1
and NQ2 are 24 and 16. For the distribution of quadrature points we
use the maps given in Ref.~\cite{Gloeckle91}. The $x''$ integration
requires typically at least 18 integration points, while for the
$\varphi''$ integration 16 points are already sufficient.  The $p$
variable is also defined in an interval $(0,p_1)\cup (p_1,p_{\rm
max})$, where $p_1$ is chosen to be 7~fm$^{-1}$ and $p_{\rm
max}=30$~fm$^{-1}$. The two intervals contain NP1 and NP2 points, and
we usually choose NP1 = NQ1 and NP2 = NQ2, respectively.

When solving Eq.~(\ref{eqinttmomexpl}) we have to carry out
two-dimensional and three-dimensional interpolations on $\hat t_{\rm
s}$ and $\hat T$.  We use the Cubic Hermitean splines of
Ref.~\cite{Hueber97a}. The functional form of those splines is
described in detail in Appendix~B of this reference and shall not be
repeated here. We find these splines very accurate in capturing the
peak structure of the two-body $t$-matrix, which occurs for off-shell
momenta $p \simeq p'$.  An additional advantage of the Cubic Hermitian
splines is their computational speed, which is an important factor,
since the integral in Eq.~(\ref{eqinttmomexpl}) requires a very large
number of interpolations.

\begin{figure}
\centerline{\psfig{file=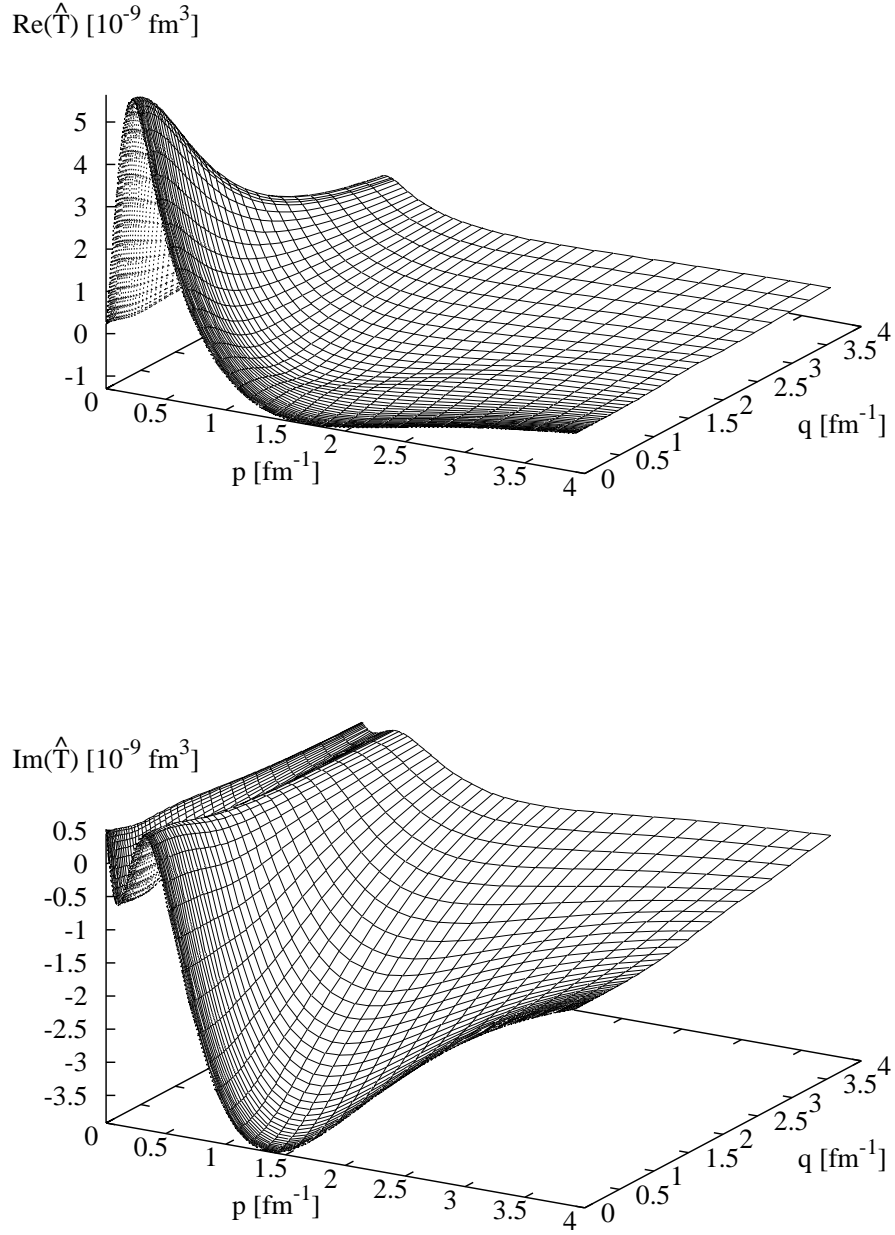,width=14cm}}
\caption {\label{figimret4} Real and imaginary part of $\hat
T(p,x_p=1,\cos \varphi_{pq} = 1,q,x_q=1,q_0)$ at 3 MeV projectile energy
as obtained from the MT-IVa potential.}
\end{figure}

\begin{figure}
\centerline{\psfig{file=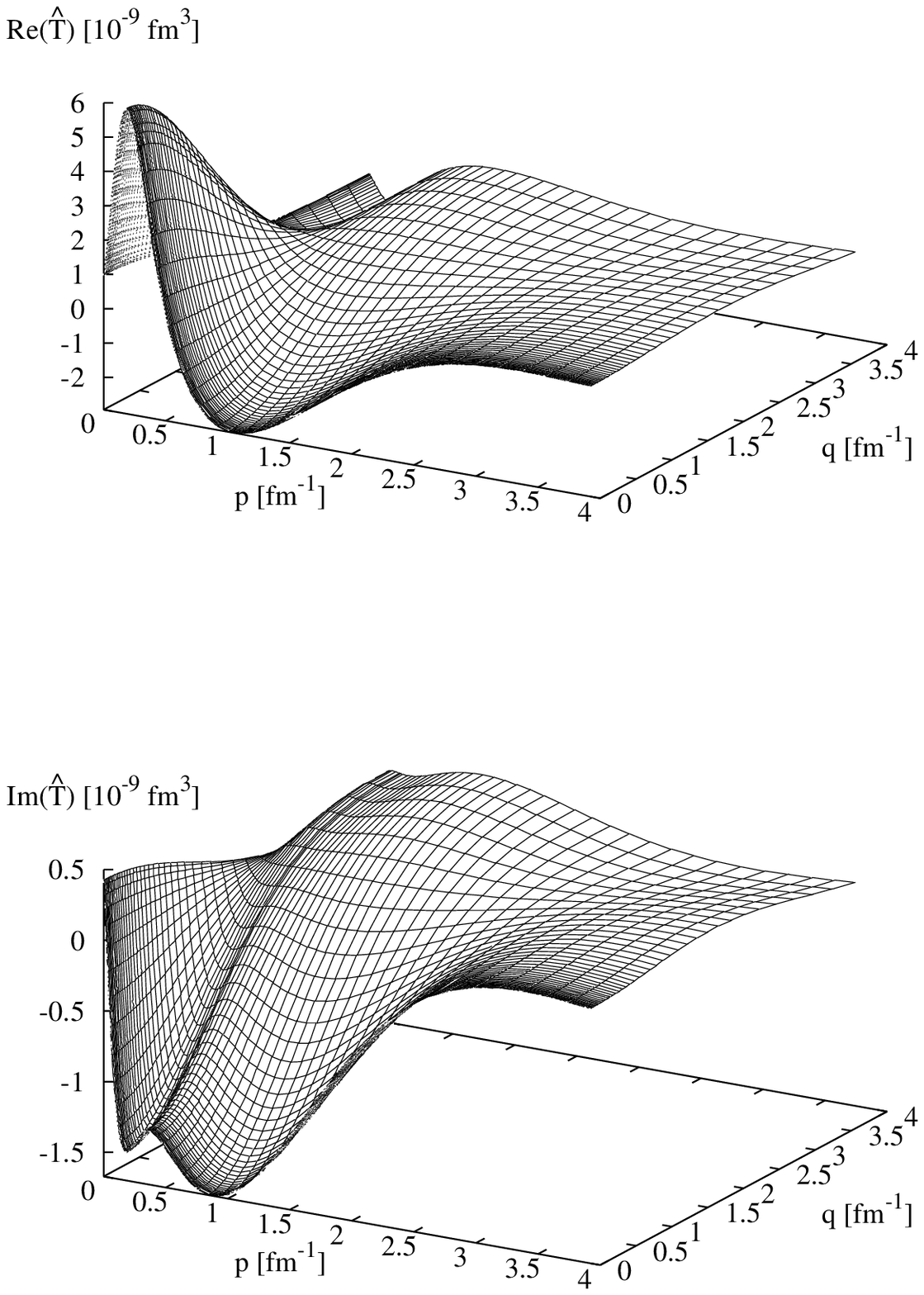,width=14cm}}
\caption {\label{figimret3} Same as Fig. 1 but for the MT-IIIa potential.}
\end{figure}

In Fig. \ref{figimret4} the real and the imaginary parts of the
scattering amplitude $\hat T(p,x_p=1,\cos \varphi_{pq} =
1,q,x_q=1,q_0)$ as obtained from MT-IVa potential are displayed. The
projectile energy is 3 MeV, and the amplitude is taken in forward
direction, i.e. the two angles are set to zero.  The figures show that
most of the structure of the amplitude is concentrated at small
momenta $p$ and $q$.  The corresponding amplitudes derived from the
MT-IIIa potential are shown in Fig. \ref{figimret3}. Though the
imaginary part has a little more structure for $p$ smaller than 1
fm$^{-1}$, the function is in general very smooth. The real real part
of $\hat T$ have for both potentials a quite similar structure.

\begin{figure}
\centerline{\psfig{file=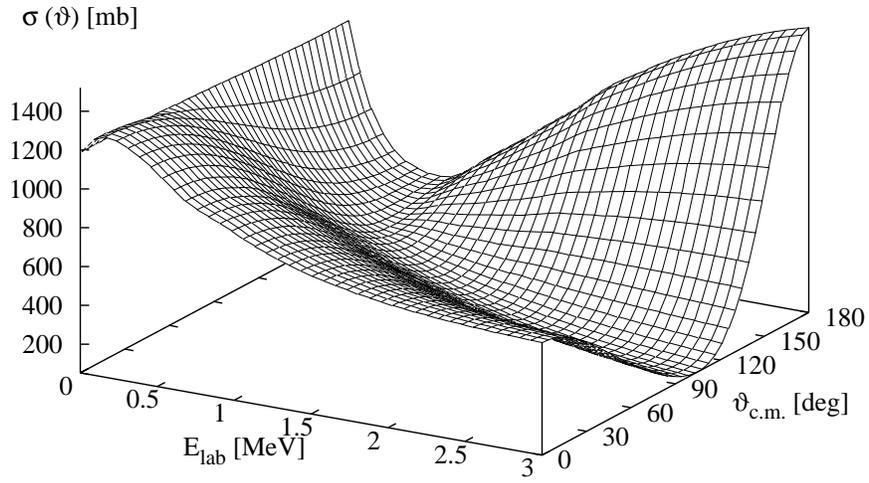,width=13.5cm,angle=-90}}
\caption {\label{figdcrmtiv}The differential cross section
$\sigma(\vartheta)$ as function of the projectile energy and
scattering angle $\vartheta$ obtained form the MT-IVa potential.}
\end{figure}

\begin{figure}
\centerline{\psfig{file=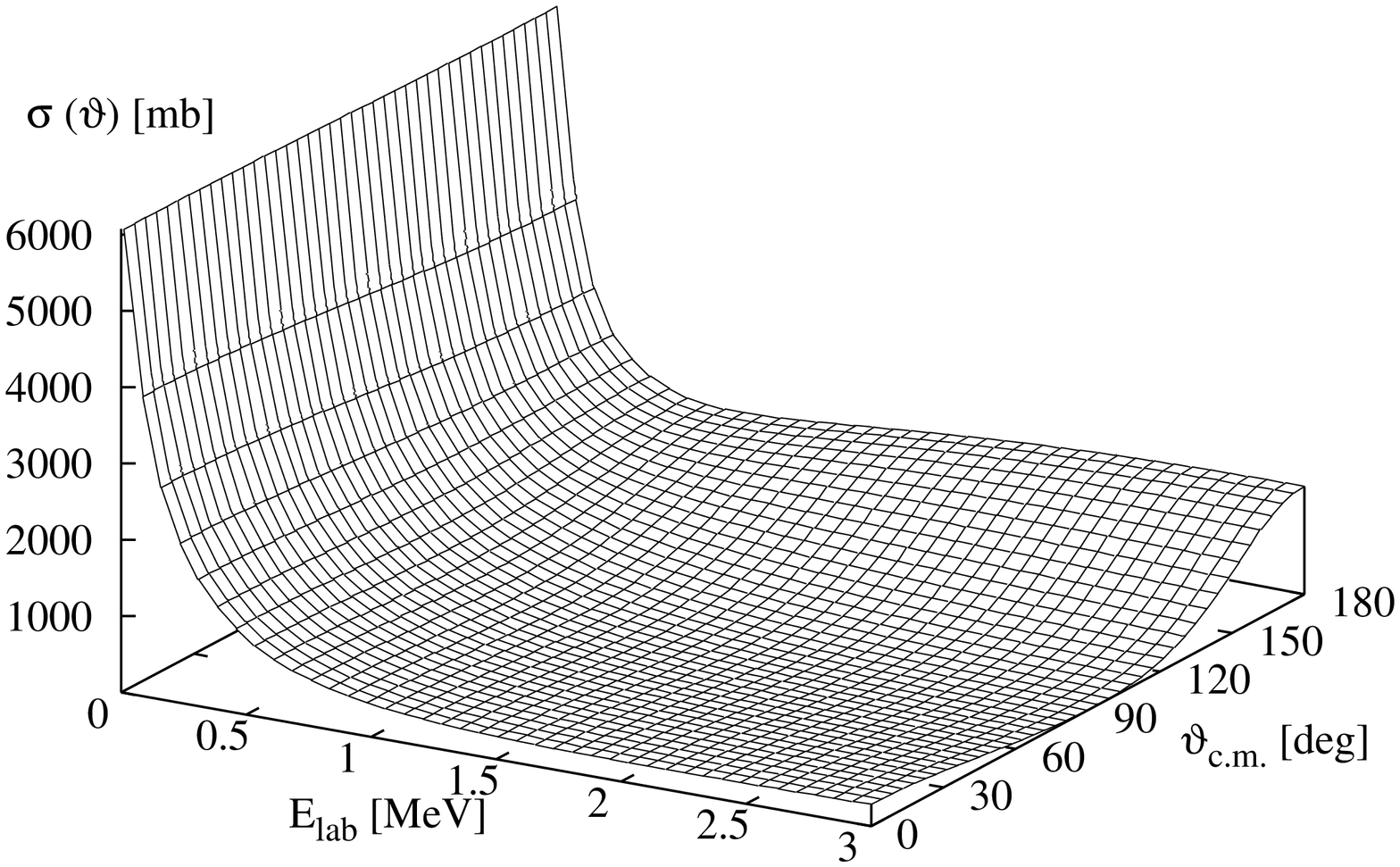,width=13.5cm,angle=-90}}
\caption {\label{figdcrmtIIIa} Same as Fig. 3 but for the MT-IIIa potential.}
\end{figure}

The solution for the transition amplitude serves as basic input to
obtain the elastic scattering amplitude according to
Eq.~(\ref{eqintumomexpl}).  In carrying out the integrals we use the
same grids as in the integral equation for $\hat T$. The differential
cross section obtained from the MT-IVa potentials is shown in
Fig. \ref{figdcrmtiv} as function of the projectile laboratory energy
$E$ and the scattering angle $\vartheta$ for energies from $0.01$ Mev
to $3.2$ MeV. As expected, for very low energies the differential
cross section is isotropic, which indicates that in a partial wave
description only $s$-waves contribute. At about 1 MeV projectile
energy the differential cross section starts to develop its more
characteristic shape, namely in forward and backward direction and a
minimum around $\vartheta = 100^{o}$. In Fig. \ref{figdcrmtIIIa} the
differential cross section obtained from the MT-IIIa potential is
displayed as function of the projectile energy and the scattering
angle. The obvious difference with respect to Fig.~\ref{figdcrmtiv} is
that the magnitude of $\sigma(\vartheta)$ obtained form the MT-IIIa is
about 5 times larger than the one obtained from MT-IVa at small
energies. The difference is related to the different values of the
3-body scattering length for the different potentials. The one for
MT-IIIa turns out to be 2.034 fm and is much larger than the one for
MT-IVa, which is 0.887 fm. These numbers are related to the different
3-body binding energies, which are -19.8625 MeV for the MT-IIIa
potential and -25.1632 MeV for the MT-IVa potential. According to
Ref. \cite{Hueber95b} the scattering length can be calculated via

\begin{equation}
a = \frac{2 \pi}{3} m \,
 \langle {\bf  q_0} \varphi_{\rm d} | U | {\bf q}_0 \varphi_{\rm d} \rangle
\bigg |_{q_0 = 0}.
\end{equation}

\begin{figure}
\centerline{\psfig{file=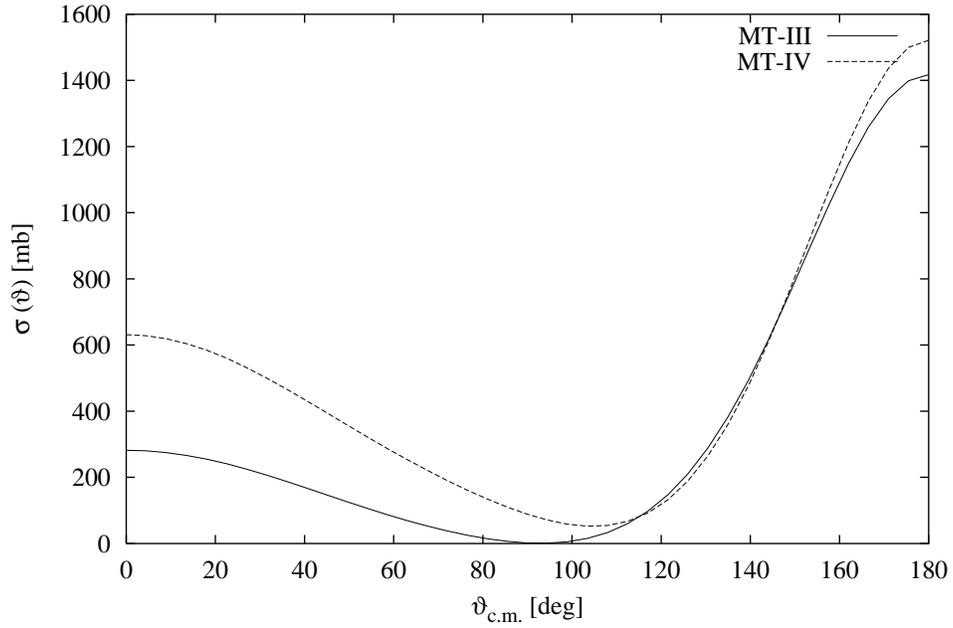,width=13.0cm,angle=-90}}
\vspace{0.5cm}
\caption {\label{figdcrmt3-mt4} The differential cross section
$\sigma(\vartheta)$ at 3 MeV projectile energy obtained from the MT-IIIa
potential (solid line) and the MT-IVa potential (dashed line).}
\end{figure}

\begin{figure}
\centerline{\psfig{file=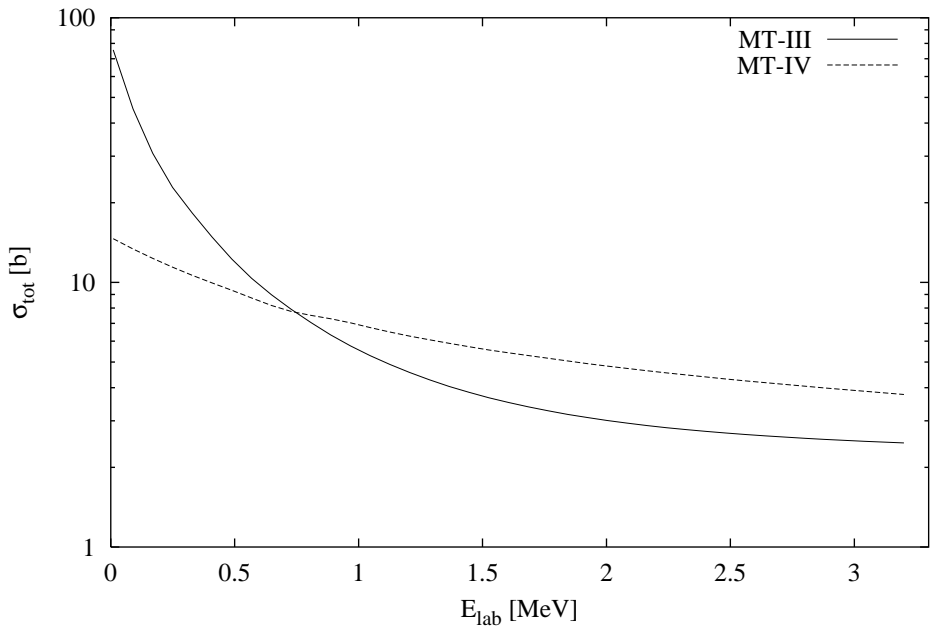,width=13.0cm,angle=-90}}
\vspace{0.5cm}
\caption {\label{figtot} The total cross section $\sigma_{\rm tot}$ as
function of the projectile energy obtained from the MT-IIIa potential
(solid line) and the MT-IVa potential (dashed line).}
\end{figure}

 Due to
the different scale the onset of a deviation from the isotropic
$\sigma(\vartheta)$ is not so easily visible in
Fig.~\ref{figdcrmtIIIa}, but it also occurs at about 1 MeV.  In order
to better compare the differential cross section obtained from both
potentials, this observable is shown in Fig.~\ref{figdcrmt3-mt4} at 3
MeV for both potentials as function of the scattering angle.  Here one
can see that $\sigma(\vartheta)$ is larger especially in forward
direction for the purely attractive potential MT-IVa. This behavior is
even clearer visible in Fig. \ref{figtot}, where the total cross
section $\sigma_{\rm tot}$ obtained from MT-IIIa and MT-IVa are shown as
function of energy.

For demonstrating and discussing the numerical stability and accuracy
of our algorithm we choose the MT-IVa potential and a fixed energy
and discuss the behavior of the observables as function of the grid
points.  Table \ref{tabnumerics} contains the total cross section
calculated at 3 MeV using Eq.~(\ref{eqdcrint}) and also via the
optical theorem, Eq.~(\ref{eqtotopt}). The first calculations listed in the
table are performed with a moderate amount of grid points in all variables.
We then successively in increase the points one variable at a time
and see that the variation in the total cross section stays about 0.3\%.
From this we conclude that our algorithm is very stable, and the numerical
error, which necessarily has to occur due to the large number of
interpolations is certainly not higher than 1\%.

We choose to consider the total cross section for this stability
study, since state-of-the-art measurements of the total cross section
have an accuracy of about 0.5\% \cite{Abfalterer98a}.  We also see the
different ways of calculating $\sigma_{\rm tot}$, differ consistently
by about 0.2\%, almost independent of the number of grid points
used. From this we can conclude, that the general error of our
calculation is 0.5\% or lower, which is for all practical purposes
sufficiently accurate. A further test of the accuracy and convergence
of our numerical calculation is in the insertion of our converged
solution for the transition amplitude $\hat T$ a further time into
Eq.~(\ref{eqinttmomexpl}) and then recalculate the observables with
this new solution for $\hat T$. The comparison is carried out for the
differential cross section at 3 MeV and listed in Table
\ref{tabcheckpade}. Here we used the results of the calculation with
the highest number of points from Table \ref{tabnumerics}, but results
for the other calculations are similar. As can be seen the agreement
of the two calculations is excellent, and we conclude that our
calculations are properly converged.

\section{A second choice of variables for the scattering equations}

In Section \ref{secscattereq}
 we described  our choice of the coordinate system used to
solve the integral equation, Eq.~(\ref{eqintt3}), for the transition
amplitude $T$. There we choose the z-axis for the integration parallel
to $\bf q$, which has the advantage of giving the free propagator in
the kernel a relatively simple functional form. It also led to
Eq.~(\ref{eqinttmomexpl}) being a three-dimensional integral equation
in 4 variables.  Obviously, the above described choice is not the only
one.  In order to test our calculations we additionally solve
Eq.~(\ref{eqintt3}) with a different choice of coordinate system. For
this specific calculation we choose the z-axis for the overall
coordinate system to be parallel to ${\bf q}_0$ as well as for the
integration in Eq.~(\ref{eqintt3}). Due to the rotational invariance of
the problem we can choose the azimuthal angle between those 2
coordinate systems to be zero. With these assumption the variables
necessary to explicitly express Eq.~(\ref{eqinttmomexpl}) are

\begin{eqnarray}
p &=& | {\bf p} |  \nonumber \\
q &=& | {\bf q} |   \nonumber  \\
x_p & = & \hat {\bf p} \cdot \hat {\bf q}_0   \nonumber  \\
x_q & = & \hat {\bf q} \cdot  \hat {\bf q}_0  \nonumber  \\
\cos \varphi_{pq} & = & \cos \varphi ({\bf p}, {\bf q})
= \hat {\bf p}_{xy} \cdot  \hat {\bf q}_{xy}  \nonumber  \\
y_q  &=& \hat {\bf q} \hat {\bf q}'' = x_q x'' + \sqrt{1 - x_q^2}
\sqrt{1 - x''^2} \cos \varphi ''   \nonumber  \\
y_p &=& \hat {\bf p} \hat {\bf q}'' = x_p x'' + \sqrt{1 - x_p^2}
\sqrt{1 - x''^2} \cos (\varphi_{pq} - \varphi '') \nonumber \\
 q'' & = & | {\bf q}'' |  \nonumber  \\
 x'' & = &  \hat {\bf q}'' \cdot  \hat {\bf q}_0  \nonumber  \\
 \cos \varphi'' & = &
\cos \varphi''({\bf q}'', {\bf q}) .
\end{eqnarray}

\noindent
The calculation of the remaining angles and momenta is straightforward
and similar to the ones given in Section \ref{secscattereq}. Thus they
are not given here. Using the above definitions of the variables we
finally arrive at the explicit expression for the transition amplitude
$\hat T$

\begin{eqnarray}
\label{eqtvarold}
\lefteqn{\langle  p, x_p, \cos \varphi_{pq},  q, x_q | \hat T | {\bf q_0}
\varphi_{\rm d} \rangle =   \varphi_{\rm d} (\sqrt{ q^2 + q q_0 x_q  +
 {\textstyle\frac{1}{4}}  q_0^2} )} \nonumber \\
  & & \;\;\; \times \,  \hat t_{\rm
s} ( p, \sqrt{\textstyle\frac{1}{4}  q^2 + q q_0 x_q + q^2_0},
\frac{\frac{1}{2} q x_{pq} + q_0 x_p}{ \sqrt{\frac{1}{4} q^2 + q q_0 x_q
+ q_0^2 }}; E - {\textstyle{\frac{3}{4m}}} q^2) \;
 \nonumber \\
& &  + \int\limits_{0}^{\infty} dq'' q''^2
\int\limits_{-1}^{1} dx'' \int\limits_{0}^{2 \pi} d\varphi''
\;  \frac{1}{E - \frac{1}{m} (q^2 + q q'' y_q + q''^2)}
 \nonumber \\
& & \,\,\, \times \,  
\hat t_{\rm
s} ( p, \sqrt{\textstyle\frac{1}{4}  q^2 + q q'' y_q + q''^2},
\frac{\frac{1}{2} q x_{pq} + q'' y_p}{ \sqrt{\frac{1}{4} q^2 + q q'' y_q
+ q''^2 }}; E - {\textstyle{\frac{3}{4m}}} q^2)
\nonumber \\
& & \,\,\, \times \,
\frac{\langle \sqrt{q^2 + q q'' y_q + \frac{1}{4} q''^2},
\frac{ q x_q + \frac{1}{2} q'' x''}{ \sqrt{ q^2 + q q'' y_q
+ \frac{1}{4} q''^2 }}
\cos \tilde \varphi, q'', x''| \hat T | {\bf q_0}
\varphi_{\rm d} \rangle}{E - \frac{3}{4m} q''^2 - E_{\rm d} + i \epsilon}.
\end{eqnarray}

As in Eq.~(\ref{eqinttmomexpl}) we solve for $\hat T$, where the
residue at the deuteron pole is explicitly taken into account as
described in Eq.~(\ref{eqtmatres}). In the form of
Eq.~(\ref{eqtvarold}) the three-body propagator has an explicit angle
dependence. In addition one has a three-dimensional integral depending
on five variables. The latter makes the numerical solution an order of
magnitude more time consuming. Thus Eq.~(\ref{eqtvarold}) is solved on
similar grids as Eq.~(\ref{eqinttmomexpl}), however with fewer
grid points.

After solving for $\hat T$, we obtain the elastic scattering amplitude by
employing Eq.~(\ref{eqintuvar}). Using the same coordinate system,
namely the $z$ axis being parallel to ${\bf q}_0$, the explicit
expression for $U$ reads

\begin{eqnarray}
\lefteqn{\langle {\bf q'} \varphi_{\rm d} | U | {\bf q_0} \varphi_{\rm d}
\rangle  =  \langle q_0 x' \varphi_{\rm d} | U | {\bf q_0}
\varphi_{\rm d} \rangle }\nonumber \\ & = & 2 \, \varphi^2_{\rm d} (q_0
\sqrt{\textstyle{\frac{5}{4}} + x'}) \left (E -
\textstyle{\frac{q^2_0}{m}} (2 + x') \right ) \nonumber \\ & & + \, 2
\int\limits_{0}^{\infty} d q'' \, q''^2 \frac{1}{E - \frac{3}{4m}
q''^2 - E_{\rm d} + i \epsilon} \int\limits_{-1}^{1} dx''
\int\limits_{0}^{2 \pi} d \varphi'' \varphi_{\rm d} (\sqrt{
\textstyle{\frac{1}{4}} q_0^2 + q_0
q'' y + q''^2 }) \nonumber \\ & & \;\;\;\;
\times \langle \sqrt{q_0^2 + q_0 q'' y + \textstyle{\frac{1}{4}}
q''^2}, \frac{q_0 x' + \frac{1}{2} q'' x''}{\sqrt{q_0^2 + q_0 q'' y+
\textstyle{\frac{1}{4}} q''^2}}, \cos \tilde \varphi, q'' , x'' | \hat
T | {\bf q}_0 \varphi_{\rm d} \rangle.
\end{eqnarray}

\noindent
Here $x' = \hat {\bf q}' \cdot \hat {\bf q}_0$ and $y = \hat {\bf q}'
\cdot \hat {\bf q}''$. The calculation of the azimuthal angle $\tilde
\varphi$ between $({\bf q} + {\textstyle{\frac{1}{2}}} {\bf q}'')$ and
${\bf q}''$ is more complicated and given in Appendix
\ref{appangle5d}.

\begin{figure}
\centerline{\psfig{file=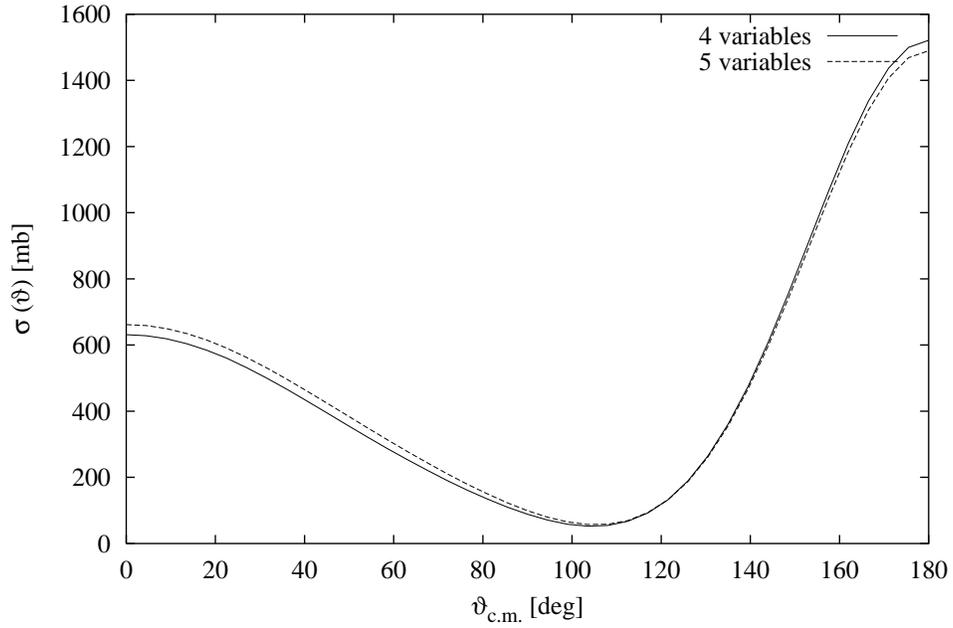,width=13.0cm,angle=-90}}
\vspace{0.5cm}
\caption {\label{figcomp}The differential cross section obtained from
the MT-IVa potential at 3.0 MeV based on the solution of $\hat T$ using
four variables (solid line) and using five variables (dashed line).
The explanation of the calculations is given in the text.}
\end{figure}

It should be clear from the beginning, that the solutions of
Eq.~(\ref{eqtvarold}) is not only much more time consuming, but will
also a priori contain a larger numerical error due to the increased
number of interpolations. In addition four dimensional interpolations
are required, whereas for the solution of Eq.~(\ref{eqinttmomexpl})
the maximum dimension for the interpolation is three.  In
Fig.~\ref{figcomp} we compare the differential cross sections at 3.0
Mev obtained from both algorithms using a medium number of grid points
only. As an aside, solving Eq.~(\ref{eqtvarold}) with the same high
number of grid points as in Eq.~(\ref{eqinttmomexpl}) is too
expensive, especially since we only had in mind to perform a rough
comparison of the two schemes. Thus, we also did not perform the same
amount of accuracy tasks as described in Section \ref{secscatterobs}
for the solution of Eq.~(\ref{eqinttmomexpl}).  As seen in
Fig. \ref{figcomp}, both solutions are reasonably close, and the
accuracy is good enough to establish, that in general both methods
give similar results. However, for practical calculations, the
procedure described in this section should not be recommended.

\section{Summary}

An alternative approach to the state-of-the-art three nucleon
scattering calculations, which are based on solving the Faddeev
equations in a partial wave basis, is to work directly with momentum
vector variables. We formulate the three-body scattering equations
below the three-body breakup threshold for identical particles as
function of vector Jacobi momenta and the projectile momentum
specifically the magnitudes of the momenta and the angles between
them. We would like to point out, that our specific formulation
and the choices of coordinate systems is also applicable above the
breakup threshold. However, here the logarithmic singularities,
inherent to the breakup, have to be treated explicitly.

As two-body force we concentrate on a superposition of an attractive
and a repulsive Yukawa interaction, which is typical for nuclear
physics, as well as on an attractive Yukawa interaction. The
corresponding two-body $t$-matrices, which enter the Faddeev equations
was also calculated as function of vector momenta. We neglected spin
degrees of freedom in all our calculations.

In order to obtain scattering observables, which are in our case the
differential and the total cross section, one solves first an integral
equation for the transition amplitude $\hat T$.  The scattering
amplitude is then obtained by an additional integration over the
half-shell amplitude $T$. This set of equations contains in essence
four vector momenta, the projectile momentum, the Jacobi momenta, and
a momentum vector as integration variable in the kernel. In principle,
one has different choices of the coordinate system, in which the
calculations are carried out. We present different choices, one
leading to a three-dimensional integral equation in four variables for
the transition amplitude $\hat T$ and one leading to a
three-dimensional integral equation in five variables for $\hat
T$. Obviously, the first choice is the preferred one. It has the
additional advantage that the free three-body propagator acquires a
relatively simple form, which will become relevant considering
scattering above the breakup threshold.

Using the transition amplitude given as function of four variables
we calculate the observables for different projectile energies and test
the accuracy and stability of our algorithms. We establish that our
calculations have an overall accuracy of less than
0.5\%, which is sufficient for all practical purposes, i.e. comparison
with experimental measurements. We also calculate the scattering observables at
one energy using the transition amplitude given as function of
five variables. The two different algorithms are in qualitative
agreement, which gives us confidence, that our calculation is correct.

Summarizing we can state that the Faddeev equations for scattering
below the breakup threshold can be handled in a straight forward
and numerically reliable fashion when using vector momenta
as variables. Our formulation allows to treat the logarithmic 
singularities above the breakup threshold in a straight forward
fashion, and work along this line is in progress.

\vfill \acknowledgements
This work was performed in part under the auspices of the U.~S.
Department of Energy under contract No. DE-FG02-93ER40756 with Ohio
University, the NATO Collaborative Research Grant 960892, and the
National Science Foundation under Grant No. INT-9726624.  We thank the
Ohio Supercomputer Center (OSC) for the use of their facilities under
Grant No.~PHS206, the National Energy Research Supercomputer Center
(NERSC) for the use of their facilities under the FY1998 Massively
Parallel Processing Access Program.

\begin{appendix}

\section{Variables for the transition amplitude $T$}
\label{anglest}

\noindent
In this case we need to rotate around the  $y = y''$ axis by the angle
 $\vartheta_q$. The vectors needed in the new coordinate system
are

\begin{equation}
{\bf Q}_0 = R(\vartheta_q) {\bf q}_0 = q_0
\left (
\begin{array}{c} - \sin \vartheta_q \\ 0 \\ \cos \vartheta_q  \end{array}
\right )
\;\;\;\;\;\;\;\;\;
{\bf Q}''
=  R(\vartheta_q) q'' = q''
\left (
\begin{array}{c}
\sin \vartheta'' \cos \varphi'' \\
\sin \vartheta'' \sin \varphi'' \\
\cos \vartheta''
\end{array} \right )
\end{equation}

\begin{equation}
{\bf P} = R(\vartheta_q) {\bf p} =
 p
\left (
\begin{array}{c}
\cos \vartheta_q \sin \vartheta_p - \sin \vartheta_q \cos \vartheta_p \\
0 \\
\sin \vartheta_q \sin \vartheta_p + \cos \vartheta_q \cos \vartheta_p
\end{array} \right ).
\end{equation}

\noindent
It is then straightforward to obtain the following angles:

\begin{eqnarray}
y_p  = \hat {\bf P} \cdot  \hat{\bf Q}''
& = & x'' x_{pq}
+ \sqrt{1-x''^2} (x_q \sqrt{1-x_p^2} - x_p \sqrt{1-x_q^2}) \cos \varphi''
\nonumber \\
y_{q_0}  =  \hat {\bf Q}''\cdot  \hat {\bf Q}_0 &= & x_q x'' - \sqrt{1-x_q^2}
\sqrt{1-x''^2} \cos \varphi'' \nonumber \\
\hat {\bf P} \cdot  \widehat{ ({\textstyle{\frac{1}{2}}} {\bf Q} +  {\bf Q}'')}
& = & \frac{ \frac{1}{2} q x_{pq} +  q'' y_p}
{\sqrt{\frac{1}{4} q^2 + q q'' x'' +  q''^2}} \nonumber \\
\hat {\bf Q}_0 \cdot  \widehat {( {\bf Q} + {\textstyle{\frac{1}{2}}}
 {\bf Q}'')}  &=& \frac{  q x_{q} +  \frac{1}{2} q'' y_{q_0}}
{\sqrt{ q^2 + q q'' x'' + \frac{1}{4} q''^2}}.
\end{eqnarray}

\section{Angles for the scattering amplitude $U$}
\label{anglesu}

We need the angle between $\hat {\bf q}_0$ and $ \hat {\bf q}''$.  It
can be calculated in terms of the integration variables in the new
coordinate system. We have chosen the $z$ axis for the integration
being in the $x$-$z$ plane of the original coordinate system, and the
azimuthal angle between them is zero. To get to the new coordinate
system we only need to rotate around $y = y''$ axis by the angle
$\vartheta'$ using the rotation matrix

\begin{equation}
R(\vartheta') =
\left(
\begin{array}{ccc}
\cos \vartheta' & 0 & - \sin \vartheta' \\
0 & 1 & 0 \\
\sin \vartheta' & 0 & \cos \vartheta' \\
\end{array}
\right).
\end{equation}

\noindent
The two vectors in the new coordinate system are given by

\begin{equation}
{\bf Q}_0 = R(\vartheta') {\bf q}_0 =
 q_0
\left (
\begin{array}{c} - \sin \vartheta' \\ 0 \\ \cos \vartheta'  \end{array}
\right )
\;\;\;\;\;\;\;\;\;
{\bf Q}'' =  R(\vartheta_q) q'' = q''
\left (
\begin{array}{c}
\sin \vartheta'' \cos \varphi'' \\
\sin \vartheta'' \sin \varphi'' \\
\cos \vartheta''
\end{array} \right ).
\end{equation}

\noindent
The angle between them is

\begin{eqnarray}
y & = & \hat {\bf Q}'' \cdot \hat {\bf Q}_0 =
-\sin \vartheta' \sin \vartheta '' \cos \varphi''
+ \cos \vartheta'  \cos \vartheta'' \nonumber \\
 & = & x' x'' - \sqrt{1-x'^2} \sqrt{1-x''^2} \cos \varphi''.
\end{eqnarray}

\noindent
The other angle we are looking for is

\begin{equation}
\hat {\bf Q}_0 \cdot  \widehat{({\bf Q}' +
{\textstyle{\frac{1}{2}}} {\bf Q}'')} = \frac{q_0 x' + \frac{1}{2} q'' y}
{\sqrt{q_0^2 + q_0 q'' x'' + \frac{1}{4}q''^2}}.
\end{equation}

\section{Angles in the five dimensional case}
\label{appangle5d}

We are looking for the azimuthal angle $\tilde \varphi$ between $({\bf
q} + {\textstyle{\frac{1}{2}}} {\bf q}'')$ and ${\bf q}''$. We had
chosen ${\bf q}$ being in the $x$-$z$ plane, now one can find the
azimuthal angle $\tilde \varphi$ by going into the $x$-$y$ plane and
using the cosine theorem

\begin{equation}
\label{eqcosine}
|{\bf q}_{xy}|^2 = |\textstyle{\frac{1}{2}}{\bf q}''_{xy}|^2
+ | ( {\bf q} + \textstyle{\frac{1}{2}}{\bf q}'')_{xy} |^2
- 2 |\textstyle{\frac{1}{2}} {\bf q}''_{xy}| \,
| ( {\bf q} + \textstyle{\frac{1}{2}} {\bf q}'')_{xy} | \cos \tilde \varphi.
\end{equation}

\noindent
The components of the vectors in the $x$-$y$ plane are given by

\begin{eqnarray}
|{\bf q}_{xy}| & = & q \sin \vartheta_q = q \, \sqrt{1 - x_q^2} \nonumber \\
|\textstyle{\frac{1}{2}}{\bf q}''_{xy}| & = &
\textstyle{\frac{1}{2}} q'' \sin \vartheta_{q''} =
\textstyle{\frac{1}{2}} q'' \sqrt{1- x''^2}  \nonumber \\
|({\bf q} + \textstyle{\frac{1}{2}} {\bf q}'')_{xy} |
&=& |  {\bf q} +  \textstyle{\frac{1}{2}} {\bf q}'' |
\sin \vartheta_{{\bf q}_0, (\frac{1}{2} {\bf q}'' + {\bf q})}
= |  {\bf q} + \textstyle{\frac{1}{2}} {\bf q}'' | \sqrt{1 - x_1^2}
\end{eqnarray}

\noindent
with

\begin{equation}
x_1 = \hat {\bf q}_0 \cdot  \widehat{( {\bf q} + \textstyle{\frac{1}{2}}
 {\bf q}'')} = \frac{\frac{1}{2} q'' x'' + q x_q}
{|  {\bf q} + \textstyle{\frac{1}{2}} {\bf q}'' |}.
\end{equation}

\noindent
Inserting these relations in Eq.~(\ref{eqcosine}) we find

\begin{equation}
\cos \tilde \varphi =
\frac{\frac{1}{4} q''^2 (1 - x''^2)
+ |   {\bf q} + \textstyle{\frac{1}{2}} {\bf q}''|^2 (1 - x_1^2)
- q^2 (1 - x_q^2)}
{q'' \sqrt{1- x''^2}
| {\bf q} +  \textstyle{\frac{1}{2}} {\bf q}'' | \sqrt{1 - x_1}},
\end{equation}

\noindent
which can be simplified to the final result

\begin{equation}
\cos \tilde \varphi =
\frac{\frac{1}{2} q'' \sqrt{1-x''^2} + q \sqrt{1-x_q^2} \cos \varphi''}
{\sqrt{\frac{1}{4} (q'' \sqrt{1-x''^2})^2 +
( q \sqrt{1-x_q^2})^2 + ( q'' \sqrt{1-x''^2})
( q \sqrt{1-x_q^2}) \cos \varphi''}}.
\end{equation}

\end{appendix}

%\bibliography{physik99b}
%\bibliographystyle{myprsty}

\clearpage

\begin{table}
\caption{ \label{tabpotpar} Parameters and  deuteron binding energy
for  the Malfliet-Tjon type potentials. As conversion factor
we use units such that $\hbar c = 197.3286$ MeV fm = 1. We also use
$\hbar^2/m = 41.47$~MeV fm$^2$.}
\begin{tabular}{l|ccccc}
\mbox{ }&$V_{\rm A}$ [MeV fm]&$\mu_{\rm A}$ [fm$^{-1}$]&$V_{\rm R}$
 [MeV fm] & $\mu_{\rm R}$ [fm$^{-1}$] & $E_{\rm d}$ [MeV] \\ \hline 
MT-IIIa & -626.8932 & 1.550 & 1438.7228 & 3.11  & -2.231   \\ 
MT-IVa & \phantom{0}-65.1776 & 0.633 & - & - & -2.223  \\
\end{tabular}
\end{table}

\begin{table}
\caption{\label{tabnumerics} The total cross section for the MT-IVa
potential at 3.0 MeV. The number of grid points for $\hat T$ are NQ =
NQ1 + NQ2, NP = NP1 + NP2, NXP, and NXQ as explained in the text. The
grid points for the integration are NX$''$ and N$\varphi''$. The grids for
the $t$-matrix are denoted with NP$_t$ and NX$_t$.}
\begin{tabular}{cccccccccc}
NQ & NP & NXP & NXQ & NX$''$ & N$\varphi''$ & NP$_t$ & NX$_t$ & $
\sigma^{\rm int.}_{\rm tot}$ &  $\sigma^{\rm opt.}_{\rm tot}$ \\
\hline
30 & 30 & 18 & 14 & 18 & 10 & 40 & 40 & 3903.18 & 3906.53 \\
30 & 30 & 18 & 14 & 18 & 16 & 40 & 40 & 3909.12 & 3913.19  \\
30 & 30 & 18 & 14 & 18 & 20 & 40 & 40 & 3909.22 & 3913.95  \\
30 & 30 & 18 & 18 & 18 & 20 & 40 & 40 & 3090.93 & 3914.09  \\
30 & 30 & 18 & 22 & 18 & 20 & 40 & 40 & 3910.06 & 3914.09  \\
30 & 30 & 22 & 22 & 18 & 20 & 40 & 40 & 3909.73 & 3914.00  \\
30 & 38 & 22 & 22 & 18 & 20 & 40 & 40 & 3909.66 & 3914.30  \\
30 & 42 & 22 & 22 & 18 & 20 & 40 & 40 & 3909.11 & 3913.45  \\
38 & 42 & 22 & 22 & 18 & 20 & 40 & 40 & 3916.66 & 3921.11  \\
42 & 42 & 22 & 22 & 18 & 20 & 40 & 40 & 3913.17 & 3917.47  \\
46 & 42 & 22 & 22 & 18 & 20 & 48 & 40 & 3911.16 & 3915.16  \\
46 & 42 & 22 & 22 & 18 & 20 & 56 & 40 & 3911.13 & 3915.02  \\
46 & 42 & 22 & 22 & 18 & 20 & 64 & 40 & 3911.11 & 3914.96  \\
50 & 42 & 22 & 22 & 18 & 20 & 64 & 40 & 3911.15 & 3915.06  \\
50 & 42 & 22 & 22 & 24 & 20 & 64 & 40 & 3910.88 & 3915.04  \\
50 & 50 & 22 & 22 & 24 & 20 & 64 & 40 & 3910.78 & 3914.93  \\
50 & 50 & 22 & 22 & 24 & 22 & 72 & 40 & 3910.79 & 3914.88  \\
50 & 50 & 22 & 22 & 24 & 22 & 80 & 40 & 3910.71 & 3914.72  \\
50 & 50 & 22 & 22 & 24 & 22 & 80 & 40 & 3910.71 & 3914.69  \\
58 & 58 & 22 & 22 & 24 & 22 & 80 & 40 & 3913.31 & 3917.22  \\
58 & 58 & 26 & 26 & 24 & 26 & 80 & 40 & 3913.95 & 3917.44  \\
58 & 58 & 30 & 30 & 24 & 26 & 80 & 40 & 3913.85 & 3917,44  \\
58 & 58 & 34 & 34 & 24 & 26 & 80 & 40 & 3913.85 & 3917.46  \\
58 & 58 & 38 & 38 & 24 & 26 & 80 & 40 & 3913.86 & 3917.61  \\
\end{tabular}
\end{table}

\clearpage

\begin{table}
\caption{\label{tabcheckpade} The differential cross section for the
MT-IVa potential at 3.0 MeV. The second column contains the cross
section obtained from the Pad{\'e} sum and the third column the cross
section obtained by reinserting the solution. The corresponding
total cross sections are 3913.86 mb and 3913.93 mb respectively.}
\begin{tabular}{ccc}
 $\vartheta$ [deg] &  $\sigma(\vartheta)$ [mb] & $\sigma(\vartheta)$ [mb] \\
\hline
     0.000       &       634.458   &      634.483 \\
     8.215       &       624.423   &      624.446 \\
    17.548       &       590.018   &      590.040 \\
    26.893       &       535.377   &      535.396 \\
    36.240       &       467.039   &      467.056 \\
    45.589       &       392.139   &      392.153 \\
    54.938       &       316.841   &      316.853 \\
    64.288       &       245.326   &      245.336 \\
    73.638       &       180.241   &      180.249 \\
    82.988       &       123.550   &      123.557 \\
    92.337       &        78.909   &       78.917 \\
   101.687       &        53.890   &       53.899 \\
   111.037       &        61.678   &       61.690 \\
   120.387       &       120.744   &      120.760 \\
   129.736       &       250.915   &      250.937 \\
   139.085       &       464.483   &      464.511 \\
   148.434       &       752.825   &      752.860 \\
   157.780       &      1074.342   &     1074.383 \\
   167.122       &      1354.729   &     1354.775 \\
   176.421       &      1508.989   &     1509.036 \\
   180.000       &      1522.727   &     1522.774
\end{tabular}
\end{table}

%%%%%%%%%%%%%%%%%%%%%%%%%%%%%%%%%%%%%%%%%%%%%%%%%%%%%%

\noindent

\noindent

\noindent

\noindent

\noindent

\end{document}